\documentclass[aps,pre,twocolumn,superscriptaddress,showpacs,preprintnumbers,longbibliography]{revtex4-1}
\usepackage{natbib}
\usepackage[english]{babel}
\usepackage[dvips]{graphics}
\usepackage{graphicx,epsfig}
\usepackage{amsmath}
\usepackage{color}
\usepackage{multirow}
\usepackage[normalem]{ulem}
\usepackage{booktabs}
\usepackage{physics}
\usepackage{dsfont}
\usepackage{float}
\usepackage{amsfonts}
\usepackage{eurosym}
\newcommand{\CB}[1]{\textcolor{blue}{#1}}

\begin{document}
\title{Optimal quantum reservoir computing for market forecasting: 
An application to fight food price crises}

\author{L. Domingo}
\email[E--mail address: ]{laia.domingo@icmat.es}
\affiliation{Departamento de Química; Universidad Autónoma de Madrid;
CANTOBLANCO - 28049 Madrid, Spain}
\affiliation{Grupo de Sistemas Complejos; Universidad Politécnica de Madrid; 28035 Madrid, Spain}
\affiliation{Instituto de Ciencias Matemáticas (ICMAT); Campus de Cantoblanco; 
Universidad Aut\'onoma de Madrid;
Nicolás Cabrera, 13-15; 28049 Madrid, Spain}

\author{M. Grande}
\email[E--mail address: ]{mgrande45@gmail.com}
\affiliation{AgrowingData; 04001 Almer\'ia (Spain)}
\affiliation{Grupo de Sistemas Complejos; Universidad Politécnica de Madrid; 28035 Madrid, Spain}

\author{G. Carlo}
\email[E--mail address: ]{carlo@tandar.cnea.gov.ar}
\affiliation{Comisi\'on Nacional de Energ\'ia At\'omica, CONICET, Departamento de F\'isica,
Av.\ del Libertador 8250, 1429 Buenos Aires, Argentina}

\author{F. Borondo}
\email[E--mail address: ]{f.borondo@uam.es}
\affiliation{Departamento de Química; Universidad Autónoma de Madrid;
CANTOBLANCO - 28049 Madrid, Spain}

\author{J. Borondo}
\email[Corresponding author E--mail address: ]{jborondo@gmail.com}
\affiliation{AgrowingData; 04001 Almer\'ia (Spain)}
\affiliation{Departamento de Gesti\'on Empresarial; Universidad Pontificia de Comillas ICADE;
 Alberto Aguilera 23; 28015 Madrid (Spain)}
\date{\today}

\begin{abstract}

The emerging technology of quantum reservoir computing (QRC) stands out in the noisy-intermediate scale 
quantum era (NISQ) for its exceptional efficiency and adaptability. 
By harnessing the power of quantum computing, it holds a great potential to untangle 
complex economic markets, as demonstrated here in an application to food price crisis prediction - 
a critical effort in combating food waste and establishing sustainable food chains. 
Nevertheless, a pivotal consideration for its success is the optimal design of the quantum reservoirs, ensuring both high performance and compatibility with current devices. 
In this paper, we provide an efficient criterion for that purpose, based on the complexity of the reservoirs.
Our results emphasize the crucial role of optimal design in the algorithm performance, 
especially in the absence of external regressor variables, showcasing the potential for novel insights 
and transformative applications in the field of time series prediction using quantum computing.

\end{abstract}

\maketitle

\section{Introduction} \label{sec.intro}

Time series forecasting plays a pivotal role in numerous disciplines, enabling informed decision-making 
and proactive planning. 
The ability to predict future trends and behaviors from historical data is essential across diverse domains, including finances, economics, weather forecasting, healthcare, and industrial processes. 
However, predictions in real-life time series pose significant challenges due to the complexities 
inherent to the data, such as the high volatility of the markets, 
and strong dependence on external factors \cite{zambrini_npj}. 

Among all the available \textit{classical} machine learning methods, reservoir computing (RC) \cite{ESN} 
has achieved notable success in time series forecasting \cite{next-genRC, ARC, RC1, timeSeries}, 
even when dealing with chaotic behavior \cite{chaos1, chaos2}. 
Moreover, the flexibility inherent to RC allows it to be implemented
using either recurrent neural networks or physical substrates, such as
silicon photonic chips \cite{physicalRC}, memristors \cite{physicalRC2}, or even
sophisticated cultured neuronal networks \cite{Yamamoto,Soriano}.
RC has shown to be extremely powerful \cite{Fujii2021,QRC2}
to forecast the complex behavior associated with market evolution \cite{anticipating}, 
something fundamental in today's world economy. 
In particular, RC can generate accurate forecasts that anticipate the future direction 
of the markets \cite{anticipating}.
In fact, for most machine learning applications in finance, such as trading, 
anticipation represents a differentiating factor. 
Moreover, in order to take full advantage of anticipation when operating in financial markets, 
the speed of the models becomes extremely important. 
This is where quantum machine learning can make the difference \cite{jacquier2022quantum}. 
Hence, there is a clear need to combine these two technologies and develop hybrid classical-quantum 
machine learning models that are feasible in NISQ hardware, such as the QRC for time series forecasting 
that is presented in this paper.

QRC emerged as a promising candidate in advancing the computational capabilities of classical RC \cite{QRC0}.
By harnessing the properties of quantum mechanics using different strategies, 
see for example Ref.~\cite{zambrini_com_phys}, QRC aims to overcome certain limitations faced 
by  classical RC, such as scalability, memory capabilities, 
and generalization to high-dimensional data, due to the exponential growth of the Hilbert space. 
Moreover, QRC perfectly adapts to the NISQ era \cite{NISQ}, 
since its simple training strategy makes it suitable for the current generation of quantum computers
\cite{Preskill18}, this fostering significant advances in quantum machine learning 
\cite{CITA, DynamicalIsing, OptQRC, Domingo_QRCNoise}.
QRC uses the dynamics of a quantum system, the \textit{quantum reservoir}, to extract useful information 
from the data, which is then fed to a classical machine learning model that predicts the desired target. 
Within this framework, QRC has been able to solve classification \cite{imageQRC},
regression \cite{CITA, quantumchemQRC,Domingo_QRCNoise}, and temporal tasks \cite{zambrini_npj, zambrini_com_phys, time_series2, time_series3,time_Series4,time_series5, QRC2}.

To ensure that the extracted features contain enough information for learning the input-output relationships,
the quantum reservoir must be a sufficiently complex quantum circuit. 
Indeed, the complexity of the random quantum circuits used as quantum reservoirs has proven to be 
crucial in achieving an optimal performance in QRC for non-temporal tasks \cite{CITA}, 
where the data points are independent of each other.
In this respect, the majorization principle \cite{majorization_original} has emerged as an efficient 
indicator for complexity of random quantum circuits \cite{majorization,benchmark}, 
because of its suitability for the NISQ era.
Other indicators, borrowed from the field of \textit{Quantum Chaos} \cite{Gu},
such as \textit{out-of-time-ordered correlators} \cite{OTOC} or Krylov complexity \cite{Krylov}, 
are also proving to be compelling ways to gauge the performance of quantum reservoirs.
Although more computationally cumbersome, they can be applied to very different substrates
\cite{Krylov_photon,Krylov_memristor,Yamamoto}, and provide valuable insight about the underlying 
physical properties of the families of quantum reservoirs \cite{OptQRC,Domingo_Krylov}.
This ability relies on the fact that these methods estimate the extent to which the information 
spreads from the input state to the reservoir.

In this paper, 
we consider the design of quantum reservoirs optimal for the more challenging case of temporal tasks.
In this respect, remarkable results have been reported for the benchmark Ising model in Refs.~\cite{DynamicalIsing,OptQRC,NatMI}.
We here show that, contrary to the Ising case, optimal performance in QRC can also be obtained 
using reservoirs composed solely of circuits featuring a much smaller number of quantum gates, 
something more in the line of the present NISQ era. 
Adapting quantum algorithms to current quantum hardware constraints is crucial, as QRC scalability 
hinges on compatibility with these limitations.
In this respect, the results of our study show that the majorization criterion provides an excellent 
method to design quantum reservoirs tailored for temporal tasks. 
Notice that this is not a mere extension of the non-temporal case, since the reservoir needs 
not only to capture information from the input quantum state encoding the dataset, 
but also make use of the memory from previous states in an efficient way.
Additionally, the reservoir can incorporate external regressor variables, 
which offer complementary insights into the time series. 
The complexity and challenges involved in this setting make it essential to obtain optimal reservoir
designs within the constraints imposed by the NISQ devices.

In particular, we here focus on analyzing the performance of QRC in predicting the 
agri-commodities market. 
Our method is able to effectively forecast the future prices of zucchini, a highly important product for agriculture in the south-east of Spain, which is very relevant both at national and European levels. 
Indeed, anticipating the evolution of the agri-commodities market is becoming crucial to alert of 
food price crises, thus contributing to the increase of sustainability \cite{Ukraine, covid,food_insecurity}. 
Regrettably, predicting in this market is far from simple and/or straightforward, since several 
difficulties have to be overcome. 
The first one arises from the fact that the time series of this market is very volatile, 
making it hard to anticipate changes in the price \textit{trends}. The second major challenge to be addressed is the interconnectivity of the market, 
where several factors such as climate, crop production, and imports/exports 
between countries can severely affect the final prices.

Two different scenarios will be considered in our study of the design and performance of the 
quantum reservoirs.
First, only the price time series is given to the model, while in the second part, 
the reservoir also makes use of external regressor variables.
Despite the mentioned challenges in the real-life case considered, the results of this paper show 
that the majorization criterion is a good performance indicator for QRC in time series forecasting.

The organization of this paper is as follows. 
In Sect.~\ref{sec:Methods}, the dataset, the studied quantum reservoirs and training details are introduced. 
The results of this study are then presented in Sect.~\ref{sec:results}. 
Finally, we conclude the paper by summarizing our main conclusions in Sect.~\ref{sec:conclusions}.

%
\section{Methods}
\label{sec:Methods}
\subsection{Dataset}
  \label{sec:dataset}
  
The case chosen to study is the prices of zucchini produced in the southeast part of Spain. 
Prices (in \euro/kg) were collected daily after the different auctions. 
Then, a more suitable time series was created by aggregating these prices weekly, 
taking the corresponding averages. 
This approach is preferred over daily predictions in the agri-food market, 
because it yields more representative results. 
In this way, a dataset spanning from March 2013 to March 2022 was built. 
For computation, this dataset is divided into three parts: training, validation, and test sets. 
The first one includes the 356 data points until December 2019, 
the second one includes 53 data points with the 2020 prices, 
and the third one includes 62 data points with prices ranging from January 2021 onward.
Due to the small number of points in the training set, obtaining accurate predictions 
without overfitting constitutes a very challenging task. 
Furthermore, the time series has been scaled to the $[0, 1]$ range using a linear scale,
to provide suitable inputs to the quantum reservoir. 
The learning process is done using the training set, while the validation set is used 
to identify the optimal hyperparameters for each model. 
Finally, the test set is used to evaluate the performance of the models on unseen data. 

In addition to the time series, our QRC algorithm can make use of regressor variables, 
that gather information on production volumes and international trade, 
this provides supplementary data to be used in the price prediction process.

\subsection{Quantum reservoir computing}
  \label{sec:QRC}

The idea behind QRC consists of using the Hilbert space, where quantum states live,
as an enhanced feature space of the input data. 
In this way, the extracted features, strengthened by quantum entangling operations, 
are used to feed a classical machine learning model, which predicts the desired target. 
For time series forecasting, as it is our case, we consider the $n-$dimensional time series $\{y(t)\}_{t}$. 
The goal is to predict the value of the time series at time $t+\Delta t$, $y(t + \Delta t)$, 
given the past values of the series $\{y(T)\}_{T\leq t}$. 
Contrary to what happens in the case of non-temporal tasks, the input series here has to be fed to 
the quantum reservoir sequentially, so that the extracted features contain information from both 
the current input and the memory from past inputs. 
This requires performing an initialization process during computation, 
which causes the quantum state of the system to become a mixed state. 
Consider then a quantum system with $N>n$ qubits. 
Then, for each time step of the training process, the state of the system is initialized as
\begin{eqnarray}
    \rho(t) &= & \ketbra{y(t)} \otimes \Tr_n\left[\rho(t-\Delta t)\right], \nonumber \\ 
    \ket{y(t)} &= & (\sqrt{1-y(t)_i}\ket{0} + \sqrt{y(t)_i}\ket{1})^{\otimes 1 \cdots n},  
    \label{eq:input}
\end{eqnarray}
where $y(t)_i$ is the $i$-th component of the time series, scaled into the $[0,1]$ domain, 
and the partial trace $\Tr_n(\cdot)$ is performed on the first $n$ qubits. 
Notice that $N>n$ is required so that the state at time $t$ has information from its past states. 
In our case, the time series is one-dimensional, i.e.,~$n=1$ and the other $N - n$ qubits are used 
to gather information from the previous values of the time series, which is captured by the partial 
trace in Eq.~(\ref{eq:input}). 
After encoding the quantum state at time $t$, the system evolves under a unitary transformation until
time $t + \Delta t$. 
The unitary transformation can be a quantum circuit, or equivalently, the evolution under a Hamiltonian. 
In the reservoir computing framework, instead of fine-tuning the dynamics of a physical system, 
the natural, inherent dynamics of the system are used to perform machine learning tasks.
Therefore, the quantum operation consists of a \textit{random} unitary operator $U$ 
sampled from a carefully selected family. The state after the unitary evolution is then
\begin{equation}
    \rho(t + \Delta t) = U \rho(t) U^\dag.
\end{equation}
At the end of each time step $t$, the expected value of some local Pauli operators 
$\{P_i\}$ are calculated, where
\begin{equation}
    \expval{P_i}(t) = \Tr \left[P_i \rho(t+\Delta t)\right].
\end{equation}
Usually, the operators $P_i$ are local $X_i$ and $Z_i$ Pauli operators applied to qubit $i$. 
Notice that, in general, a $N$-qubit unitary $U$ transforms a simple observable $Z$ (or $X$)
into a linear combination of Pauli operators
\begin{equation}
    U Z_1(t)U^\dag = \sum_i \alpha_i P_i(t),
\end{equation}
where $\{P_i\}$ are tensor products of local Pauli operators. 
Therefore, measuring single Pauli operators of a state which has received a unitary transformation 
could produce complex nonlinear outputs, which could be represented as a linear combination of 
exponentially many nonlinear functions \cite{QRC}. 
The operator $U$ is chosen to create enough entanglement to generate useful features 
from the input data, while being experimentally feasible at the same time. 
Finally, after a sufficiently long training time, the extracted features 
$\hat{x}(t)=\{\expval{P_i}(t)\}=(\expval{X_1}(t),\expval{Z_1}(t),\cdots\expval{X_N}(t),\expval{Z_N}(t))$ 
are fed to a classical machine learning algorithm which predicts the target $y(t + \Delta t)$. 
Even though complex machine learning models can be used, the quantum reservoir should be able to 
extract valuable features, so that a simple machine learning model can predict the target 
$y(t + \Delta t)$. 
Usually, a linear model with regularization, such as the ridge regression, is enough to learn the output. 
Ridge regression minimizes the following expression
\begin{equation}
   MSE_r(\hat{y}, y) = \frac{1}{t} \sum_{t=1}^T \left[(y(t) - W \hat{x}(t)\right]^2 
       + \gamma \; ||W||^2 ,
 \label{eq:ridge}
\end{equation}
where $\gamma$ is the regularization parameter and $W$ are the linear coefficients. 
This loss function prevents the algorithm from learning too big coefficients $W$, 
which usually leads to unstable training and poor generalization capacity. 
In Eq.~(\ref{eq:ridge}) $\gamma$ is a hyperparameter, 
which needs to be tuned depending on the problem at hand. 
When $\gamma$ is too large, the model will learn very small values of $W$, which leads to predicting 
constant values. 
On the other hand, if $\gamma$ is too small chances of overfitting increase. 
Moreover, when the number of qubits of the system is small, the size of the extracted features containing 
the expectation values of the local Pauli operators may be quite small.
For this reason, one can gather the readout signals from intermediate times between $[t, t+ \Delta t]$. 
In this case, the input timescale $\Delta t$ is divided into $N_v$ time steps, such that the intermediate 
steps are given by $t^k = t + k \Delta t/N_v$,  $k=0, \cdots, N_v$. 
Then, the feature vector $\hat{x}(t)$ contains the 2$N$ expected values from the $N$ qubits 
and the intermediate times $t^k$, leading to a total dimension of $2N\cdot N_v$. 
Additionally, features can be expanded by adding extra regressor variables containing useful 
external information. 
In this case, the feature vector is $\hat{x}_{\text{extended}}(t) = (\hat{x}(t), x_{\text{extra}}(t))$. 
In our case, we take $N_v=2$ and $N = 7$ qubits (see the reason for this choice in the discussion 
related to Fig.~\ref{fig:num_qubits} below), where $n=1$ qubit is used to encode the input series. 
The regularization parameter is set to $\gamma = 1 \cdot 10^{-10}$ when the feature vector 
\textit{does not} include the external factors [$\hat{x}(t)$], and $\gamma = 0.1$ 
when the feature vector \textit{does} include the external factors [$\hat{x}_{\text{extended}}(t)$]. 
Two scenarios are considered in this work (see Sect.~\ref{sec:results}): 
with and without the use of regressor variables in the prediction.

\subsection{Families of quantum reservoirs}
  \label{sec:QR}
\begin{figure*}
\includegraphics[width=0.95\textwidth]{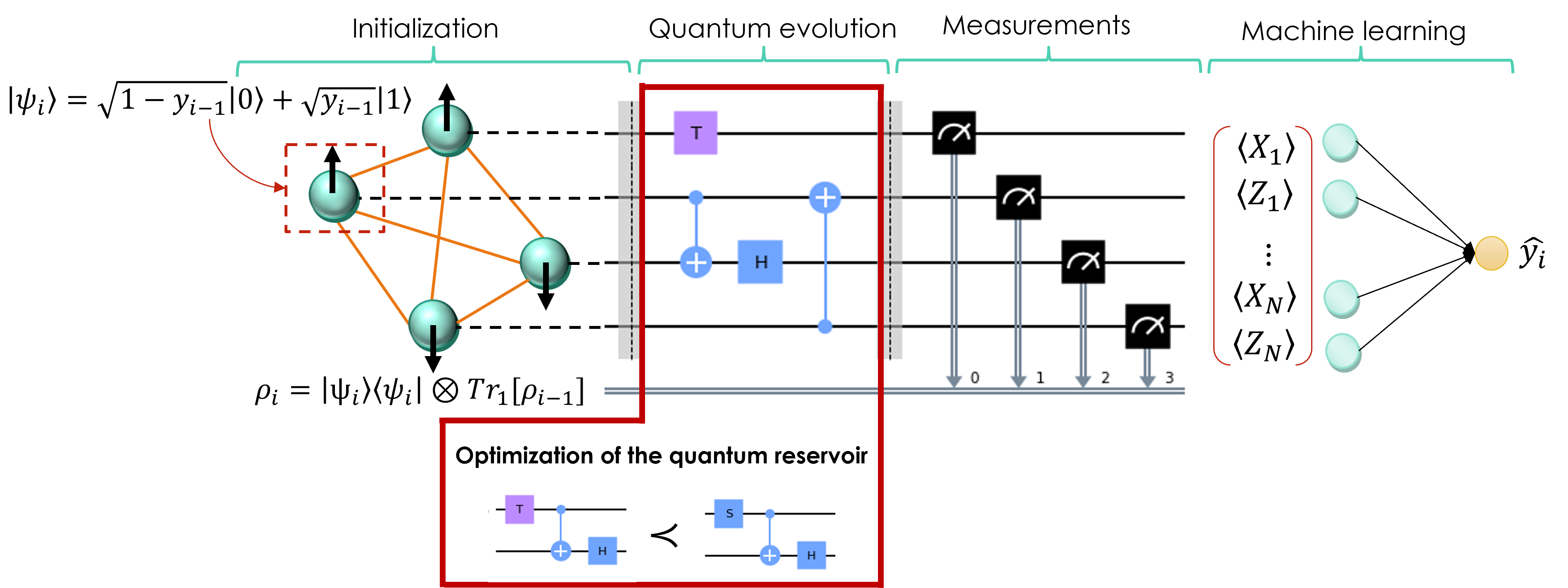}
\caption{Pipeline used to train the quantum reservoir computing model. 
The time series at step $t-1$ is given as an input. 
Then, the reservoir evolves by applying a random quantum circuit sampled from one 
of the seven families studied in this work.
Local Pauli operators are then measured and fed to the classical machine learning algorithm, 
a linear regression model. 
The choice of the quantum reservoir is optimized according to the majorization 
principle introduced in Ref.~\cite{majorization}.}
\label{fig:6}
\end{figure*}

The design of the quantum reservoir is crucial for the performance of the algorithm. 
For this reason, seven \textit{families} of quantum circuits, with different degrees of complexity 
according to the majorization principle~\cite{majorization}, have been studied. 
For each family, the quantum circuit is built by adding a fixed number of random quantum gates 
taken from such family. 
For each family, 100 simulations were carried out. 
The setting of the QRC algorithm is illustrated schematically in Fig.~\ref{fig:6}. 
The seven families of quantum reservoirs are described as follows
\begin{enumerate}
    \item \textbf{G1:} The quantum circuit is constructed from the generator G1 = \{CNOT, H, X\}, 
    where CNOT is the controlled-NOT gate, H stands for Hadamard, and X is the NOT gate. 
    Set G1 generates a subgroup of the Clifford \cite{G1} group, and thus is non-universal
    and classically simulatable.
    \item \textbf{G2:} The quantum circuit is constructed from the generator G2 = \{CNOT, H, S\}, 
    being S the $\pi/4$ phase gate. Circuits constructed from G2 generate the whole 
    Clifford group \cite{G2}, so they are non-universal and classically simulatable, 
    but more complex than G1 circuits.
    \item \textbf{G3:} The quantum circuit is constructed from the generator 
    G3=\{CNOT, H, T\}, where $T$ is the $\pi/8$ phase gate. G3 is universal and thus 
    approximates the full unitary group $U(N)$ to arbitrary precision.
    \item \textbf{Matchgates (MG):} Two-qubit gates formed by 2 one-qubit gates, 
    $A$ and $B$, with the same determinant. $A$ acts on the subspace spanned by $\ket{00}$ 
    and $\ket{11}$, while $B$ acts on the subspace  spanned by $\ket{01}$ and $\ket{10}$. 
    $A$ and $B$ are randomly sampled from the unitary group U(2):
    \begin{equation}
        G(A,B) = 
        \begin{pmatrix} 
            a_1 & 0 & 0 & a_2 \\
            0 & b_1 & b_2 & 0 \\
            0 & b_3 & b_4 & 0 \\
            a_3 & 0 & 0 & a_4
        \end{pmatrix}, \quad |A| = |B|.
    \end{equation}
    Matchgates circuits are also universal (except when acting only on nearest neighboring lines) \cite{matchgates1, matchgates2}. 
    \item \textbf{Diagonal-gate circuits ($\mathbf{D_2}$, $\mathbf{D_3}$, $\mathbf{D_N}$):} 
    The last families of gates considered are diagonal in the computational basis. 
    The diagonal gates are separated into 3 families: $D_2$, $D_3$ and $D_N$. 
    Here, $D_2$ gates are applied to pairs of qubits, $D_3$ gates are applied to 3 qubits, 
    and $D_N$ gates are applied to all qubits,
    \begin{equation}
        D_k(\phi_1, \cdots, \phi_{2^k}) = 
        \begin{pmatrix} 
            e^{i\phi_1} & 0 & \cdots & 0  \\
            0 & e^{i\phi_2} & \cdots & 0 \\
            \vdots & \vdots  & \ddots & \vdots \\
             0 & 0 & \cdots & e^{i\phi_{2^k}}
        \end{pmatrix}, 
    \end{equation}
    for $k \in \{2,3,N\}$, and with $\phi_i$ chosen uniformly from $[0, 2\pi) \ \forall i$. 
    The gates are applied on all combinations of $k$ (out of $N$) qubits, the ordering being random. 
    At the beginning and at the end of the circuit (after the initialization of the state), 
    Hadamard gates are applied to all qubits. As diagonal gates commute, they can be applied simultaneously.
    Diagonal circuits cannot perform universal computation but they are not always classically 
    simulatable \cite{diagonals}. 
    As opposed to the other families of circuits, which can be of arbitrary depth, the diagonal 
    $D_2$, $D_3$ and $D_N$ families contain a fixed number of gates, being those 
    $\binom{N}{2}$, $\binom{N}{3}$ and 1 gates, respectively. 
\end{enumerate}
\begin{figure*}[t!]
    \centering
    \includegraphics[width=0.95\textwidth]{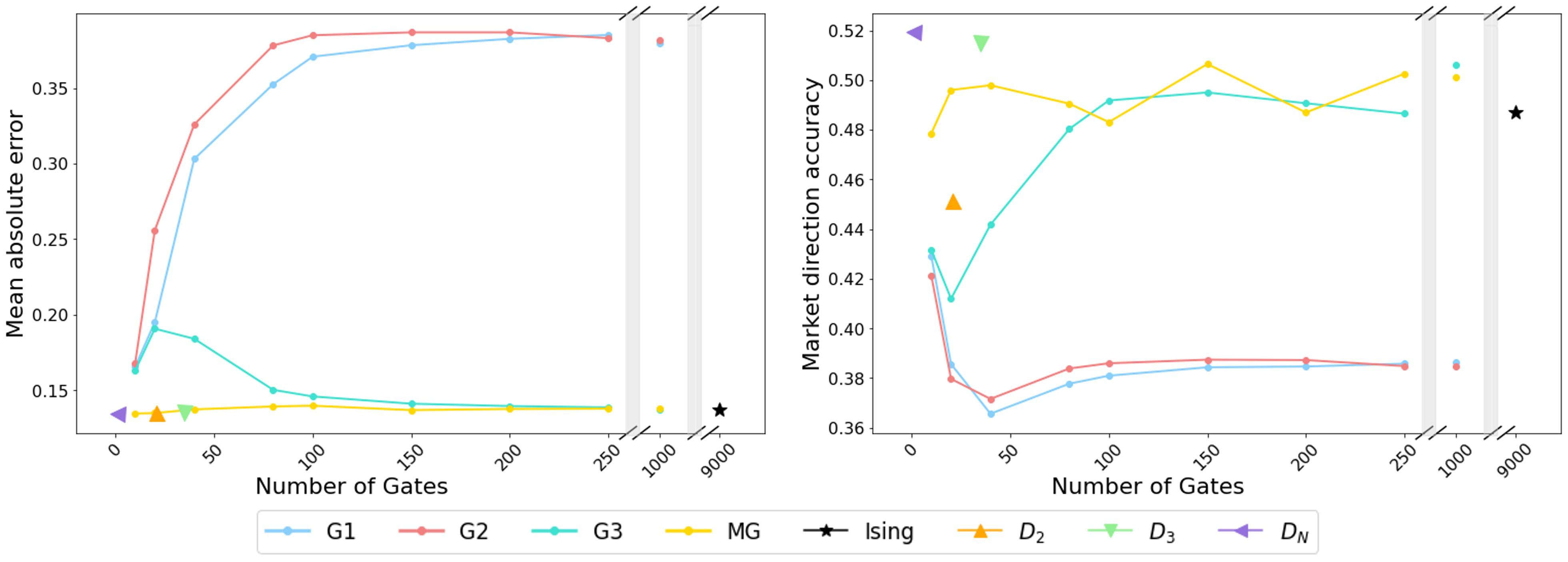}
    \caption{Average mean squared error and market direction accuracy of the eight families 
    of quantum reservoirs as a function of the number of gates of the circuit. 
  Averages are made over 100 simulations. 
  The machine learning task consists of forecasting the zucchini prices \textit{without} 
  using external regressor variables.}
  \label{fig:no_vars}
\end{figure*}

These families can be ordered in terms of complexity according to the majorization principle in Ref.~\cite{majorization}. The G3 family has the highest complexity, followed closely by MG. 
This agrees with the fact that both families are universal, and thus approximate the whole space of operators.
The diagonal circuits $D_3$ and $D_n$ are next in terms of complexity. 
The $D_2$ family has a slightly smaller complexity. Finally, the G2 and G1 are the less complex families 
of circuits, the complexity of G2 being slightly higher than the complexity of G1. 
Again, this agrees with the two families not being universal, and the G1 being a subgroup of the 
Clifford group. 

Additionally, we compare in our work the results from studied families with those obtained with 
the widespread Ising model, which has been very often used as a quantum reservoir 
\cite{Fujii2021, QRC2, quantumchemQRC, OptQRC, DynamicalIsing}. 
The Ising Hamiltonian is a model for a many-body system which has an exact analytical solution \cite{Ising}. 
It describes a spin-1/2 chain with nearest-neighbor interaction in an external magnetic field. 
In gate-based quantum computing, the quantum circuit performs the time evolution of a quantum
state under the random transverse-field Ising Hamiltonian
\begin{equation}
    H_{\text{Ising}} = \sum_{i,j=0}^{N-1} J_{ij} Z_iZ_j + \sum_{i=0}^{N-1} h_{i} X_i,
    \label{eq:Ising}
\end{equation}
where $X_i$ and $Z_j$ are Pauli operators acting on the site $i, j$-th qubit. 
The parameters are chosen here in the same way as in Ref.~\cite{DynamicalIsing} 
since this method provides a state-of-the-art method to select optimal parameters of the Ising model for QRC.
In this case, $J_{ij}$ are sampled from the uniform distribution $U(-J_s/2, J_s/2)$ and $h_i = h$ are constant.
The optimal parameters in Ref.~\cite{DynamicalIsing} fulfill $h/J_s = 0.1$. 
All time evolutions are performed for a lapse of time $T =10$.  

\section{Results}
  \label{sec:results}

\begin{figure*}[t]
    \centering
    \includegraphics[width=0.85\textwidth]{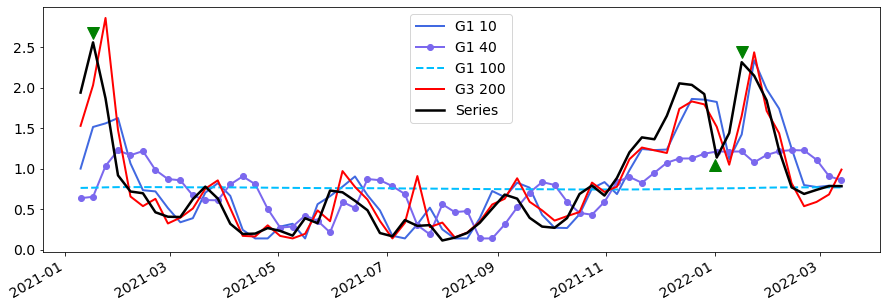}
    \caption{Prediction of the zucchini price (in euros/kg) time evolution in the 
    test set 2021-2022 campaign for the G1 family with 10, 40 and 100 gates 
    \textit{without} regressor variables, compared with that for the G3 family 
    with 200 gates \textit{with} regressor variables, 
    and the actual price series in black.}
  \label{fig:example_pred}
\end{figure*}

\subsection{Predicting without regressor variables} \label{sec:scenario1}

Let us consider the first scenario, in which only the time series is provided to the quantum reservoir, 
and no other (external) factors are considered in the prediction. 

Figure~\ref{fig:no_vars} shows the performance in predicting the time series of zucchini 
prices for the eight families of quantum reservoirs used here (see description in Sect.~\ref{sec:QR})
as a function of the number of gates used in the circuit. 
The performance of the method is evaluated using two metrics: 
the standard Mean Absolute Error (MAE) (left panel), and the more interesting and informative 
Market Direction Accuracy (MDA)  (right panel), introduced by the authors in Ref.~\cite{anticipating}. 
In this respect, it should be remarked that our MDA effectively checks whether a model can accurately 
predict sudden changes in the trend of a time series, something that is critical for the kind of data 
considered in this work (and others). 
Notice that low values of the MAE or high values of MDA are indicative of good predictions. 
In Fig.~\ref{fig:no_vars}, we also plot the asymptotic values of the MAE and MDA, 
evaluated with a large number of gates, i.e.~1000, and the results obtained
with the Ising model, which requires implementing a circuit with a large number, i.e.~9000, of gates. 
As can be seen, our results clearly indicate that there are significant differences 
in the performance calculated with the different gate families.
In particular, the worst results, both in terms of MDA and MAE, are obtained when the G1 and G2 circuits
are used. 
This is consistent with the fact that these families of quantum circuits are less complex, 
according to the  majorization criterion \cite{majorization}, since they are not universal 
and generate unitaries within the Clifford group \cite{CITA}. 
Furthermore, as the number of gates increases, the performance of the QRC algorithm worsens in this case.
Indeed, when the number of gates exceeds approximately 100, the MDA drops to around 38\%, 
which is equivalent to a random guess, and the MAE also reaches up to a value equivalent 
to a random prediction. 

This trend is also evident in Fig.~\ref{fig:example_pred}, which depicts the prediction 
in the test set of the G1 family for 10, 40, and 100 quantum gates. 
As the number of gates increases, the variability of the predictions decreases, 
eventually converging to the extreme case where the predictions are just a constant value similar 
to the average value of the time series (indicated with a light-blue dashed line). 
This implies that the quantum reservoirs generated with the G1 (and also the G2) families, 
instead of extracting useful features from the input data, slowly suppress the information encoded 
in the quantum states.
This behavior worsens as the number of gates from these families increases, 
reaching the worst-case scenario at around 100 gates, where the information 
about the input series is completely lost.

The opposite behavior is exhibited by the G3 family; see results in green in Fig.~\ref{fig:no_vars}. 
In this case, the performance of the QRC algorithm improves with the number of quantum gates, 
converging to the optimal value of MAE=0.13 and MDA=0.5 for around 150 quantum gates. 
The optimality of the G3 family performance agrees well with the fact that it has the highest 
complexity, according to the majorization criterion \cite{majorization}. 
Moreover, the predictions obtained with the G3 family and the aid of external regressor variables 
are also shown in Fig.~\ref{fig:example_pred} for comparison. 
These results will be discussed in Sect.~\ref{sec:scenario2} below. 

The results in Fig.~\ref{fig:no_vars} also indicate that the performance of the MG circuits 
is very similar to that of the G3 family, both in terms of MAE and MDA.
However, the optimal error is achieved with only 40 quantum gates, a value which is significantly lower than the number of gates required by the G3 family. 
Despite this advantage, the implementation of the MG quantum gates is much more complicated than 
that for the elementary quantum operations required for the G3 family. 
Therefore, although both the MG and G3 circuits perform equally well, 
the G3 circuits are more suitable for NISQ devices. 
Furthermore, the MG family is slightly less complex than the G3 family in terms of the majorization criterion,
but this does not seem to have any significant impact on the performance of the QRC algorithm.

Regarding the diagonal circuits, their performance is very similar to the optimal performance,
as seen in the results of Fig.~\ref{fig:no_vars}. 
However, the $D_2$ circuits exhibit significantly lower values of MDA, which also agrees with the fact 
that they are less complex according to the majorization criterion. Notice that the diagonal circuits have a fixed number of gates, and therefore only one value of MSE and MDA is shown in Fig.~\ref{fig:no_vars}.

Another important point is the performance of the quantum reservoir generated by the
time evolution under the Ising model, and how it compares with our optimal G3 results.
The corresponding results are shown in Fig.~\ref{fig:no_vars}, and 
as can be seen, the performance is also optimal both in terms of MAE and MDA.
In Refs.~\cite{CITA,Domingo_QRCNoise}, a method to estimate the number of gates required 
to implement the Ising model on a gate-based quantum computer with gates sampled only 
from the G3 family was discussed. 
Since not all quantum gates can always be implemented in current quantum devices with perfect accuracy, 
the rotations present in the implementation of the Ising model are usually decomposed in $H$ and $T$ gates, leading to an implementation of the Ising model using gates from the G3 family. 
Consequently, it is worth comparing here the number of G3 gates required to implement the Ising model 
with the number of G3 gates needed to achieve optimal performance.
We conducted 100 simulations varying the parameters $J_{ij}$ and $h_i$ of the Ising Hamiltonian 
(see Sect.~\ref{sec:QR} below). 
The results are presented in Fig.~\ref{fig:Ising}, which shows the probability distribution 
of the number of G3 gates needed to implement each of the 100 simulations. 
The orange curve shows the best fit to the empirical distribution, where the average number 
of gates (vertical red line) appears to be very close to 9000.
This value is significantly larger than the number of gates required for an optimal 
quantum reservoir using the G3 family. 
Therefore, we conclude that implementing a time evolution operator based on the Ising model using 
gates from the G3 family is far more inefficient in terms of the number of
gates than those required to obtain an optimal quantum reservoir.

\begin{figure}[b!]
    \centering
    \includegraphics[width=0.95\columnwidth]{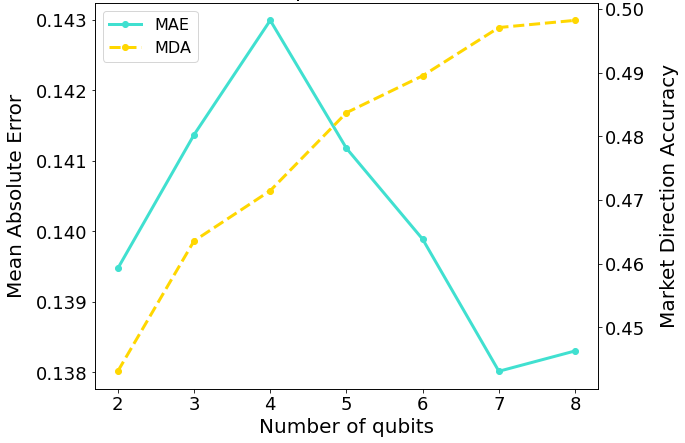}
    \caption{Performance of the G3 family, evaluated in the test set at forecasting
    the zucchini prices \textit{without} using external regressor variables with 150 
    quantum gates as a function of the number of qubits $N$. 
    Averages are made over 100 simulations.
    }
  \label{fig:num_qubits}
\end{figure}

To conclude this part, we consider the effect of the number of qubits $N$ on the performance of QRC. 
For this purpose, we have performed 100 simulations using the optimal quantum reservoirs, i.e.,
the G3 reservoirs with 150 gates (see Fig.~\ref{fig:no_vars}), for $N=2,3,\cdots, 8$ qubits.
The results are shown in Fig.~\ref{fig:num_qubits}. 
As can be seen, the performance of QRC improves in general with the number of qubits, 
until it converges to the optimal performance at $N=7$. 
For larger values of $N$, more information from the previous values of the time series is given 
to the quantum reservoir, thus yielding better results. 
However, increasing the number of qubits also increases the computational complexity of the task. 
Our numerical experiments show that, in this case, $N=7$ qubits is the appropriate balance 
between computational complexity and performance in the machine learning task.

\begin{figure}
    \centering
    \includegraphics[width=0.99\columnwidth]{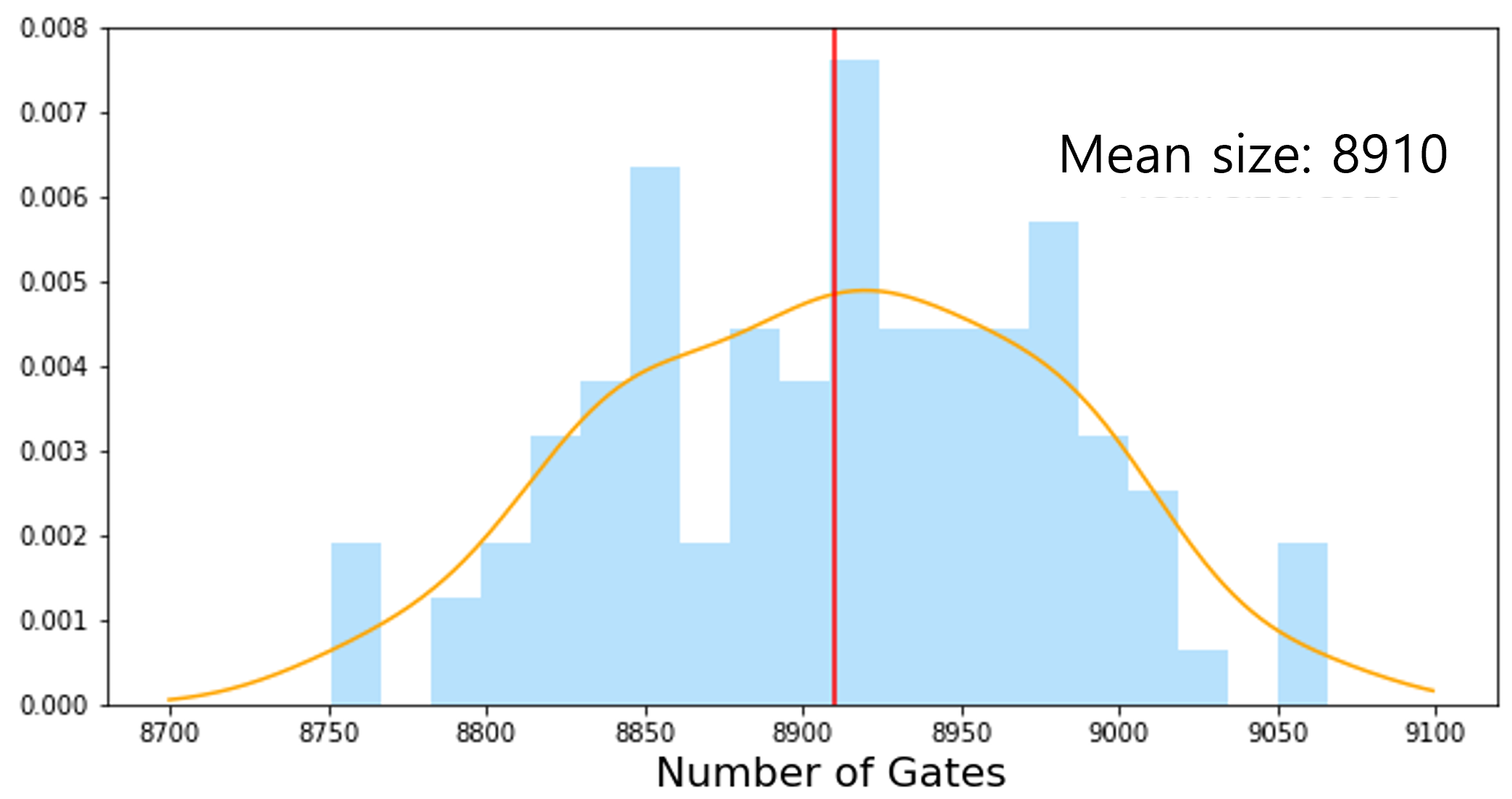}
    \caption{Probability distribution of the number of gates required to implement an Ising model in a quantum circuit using the gates from the G3 family. The y-axis shows the probability density as a function of the number of gates of the circuit. The orange curve represents the best fit of the empirical probability density function, whose average is marked with the red line.}
    \label{fig:Ising}
\end{figure}

\begin{figure*}
    \centering
    \includegraphics[width=0.95\textwidth]{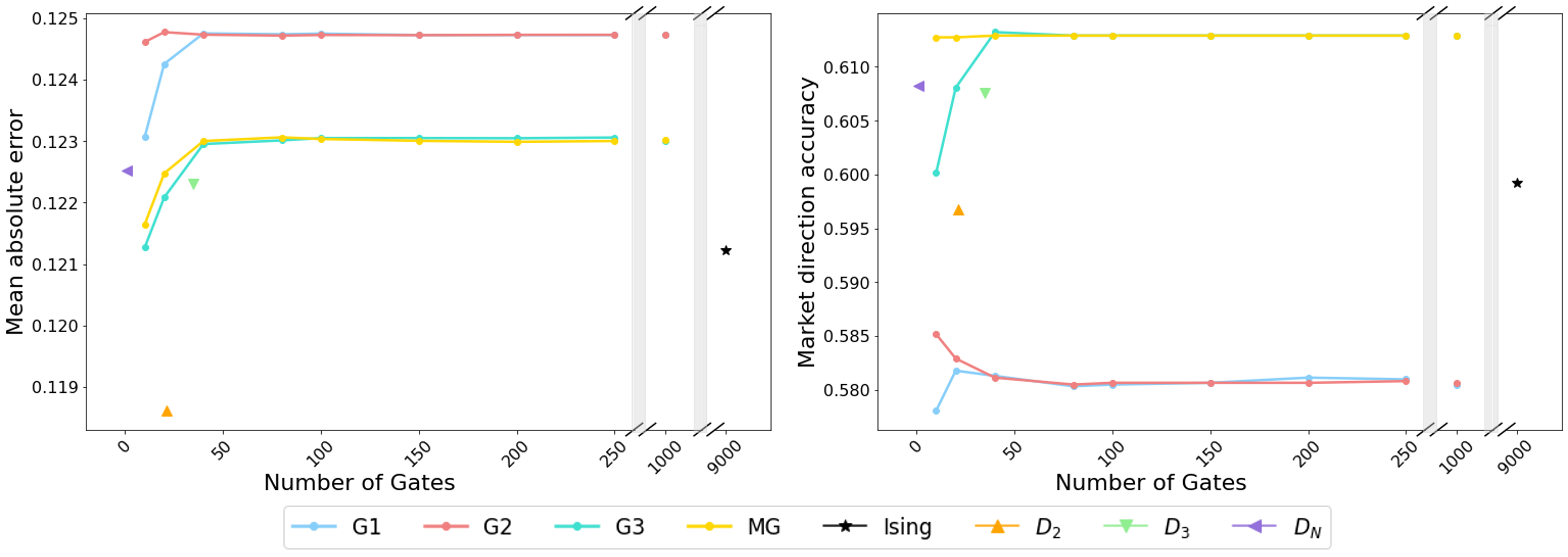}
    \caption{Same as Fig.~\ref{fig:no_vars} 
    using the external regressor variables described in Sec.~\ref{sec:dataset}.}
    \label{fig:vars}
\end{figure*}

\subsection{Getting help from regressor variables} \label{sec:scenario2}

Let us consider now the second scenario, in which external regressor variables are considered.
These variables are known for providing highly relevant information to the behavior of  
financial time series. 
For that purpose, the previous analysis was repeated but increasing 
the feature vector with nine external variables (see Sect.~\ref{sec:dataset} below). 
The performance of the best model, namely the G3 family with 200 quantum gates, is depicted in Fig.~\ref{fig:example_pred}, where it is compared with the worst model, the G1 family with 100 
gates and \textit{no} regressor variables. 
As can be seen, the new model is much better at predicting changes of tendency in the data, 
even though some abrupt ones, marked with green triangles, are not properly predicted 
due to the high complexity of the task.

The performance of the eight families of quantum reservoirs described in Sect.~\ref{sec:QR} 
are shown in Fig.~\ref{fig:vars} (similar to Fig.~\ref{fig:no_vars} in the first scenario). 
The first thing to notice is that the differences in the performance of the models 
with gates G1 \textit{vs.~}G3 are now significantly smaller. 
Indeed, in the first scenario considered before, the difference in MAE between the G1 and G3 families 
was typically around 0.23~\euro/kg, while the same magnitude when considering the external variables 
amounts only now to just 0.003~\euro/kg. 
Similarly, in terms of MDA, the difference between the G3 and G1 families without considering the 
regressor variables was $13\%$, while the same difference drops to $3\%$ when including the influence 
of the external factors.
These results indicate that the external factors contain critical information for the predictive model,
which is not related to the design of the quantum reservoir. 
Despite the strong influence of external variables considered, the difference in performance based on the majorization criterion is still apparent, something which confirms the significant dependency 
of the QRC algorithm with the reservoir design. 

Regarding the other quantum reservoir types, we can say that the MG families still provide optimal performance, 
while converging with slightly fewer gates than the G3 family. 
The $D_3$ and $D_n$ circuits provide similar MAE and slightly worse MDA than the G3 family, 
which agrees with the fact that they have somewhat smaller complexity, according to the majorization criterion.
The $D_2$ family presents a significantly worse MDA, which lies in between the G1 and G3 MDA values. 
This effect was also obtained in the first scenario and also agrees with the prescription of the 
majorization indicator. 
However, the $D_2$ circuit presents the lowest MAE, even though it is just 0.004~\euro/kg lower than 
the G3 MAE. 
This can be explained by noting that models whose predictions just tend to copy the previous values
of the series \cite{anticipating} will present lower MAE but high MDA, thus providing worse overall performance. 
For this reason, we believe that the G3 family should be considered to render better results than the $D_2$.

Finally, the Ising model exhibits similar behavior as the $D_2$ family, where the MAE is smaller than 
that of the G3 family, but the MDA is also smaller than the optimal one. 
For this reason, in this case, we conclude that when regressor variables are considered the G3 family presents optimal performance in forecasting the price series, 
and remarkably this optimal value is obtained with only 60 quantum gates.

\section{Conclusions} \label{sec:conclusions}
In the recent stream of growing interest in developing quantum methods for NISQ devices, 
aimed to outperform classical methods using the quantum computers currently available, 
the recently introduced QRC algorithm undoubtedly stands out.
Among the reasons for this success, its easy training strategy and suitability of implementation in NISQ devices should be mentioned as the most important.
The QRC algorithm uses the inherent dynamics of a quantum system, 
known as the quantum reservoir, to uncover the patterns hidden in the input data, 
which are then cranked by a classical machine learning model.
At this point, it should be remarked that the design of the quantum reservoir is crucial to guarantee 
an optimal feature extraction process.
Moreover, this is an issue of uttermost importance when dealing with tasks involving time series
(as opposed to non-temporal problems), 
where the complexity grows by having to consider not only the relation between the current input and output,
but also relationships with previous inputs. 

In this paper, we have examined the relation between quantum reservoir design and its ability to 
predict a challenging agri-commodity price time series. 
Such series, as well as others in the financial sector, are especially hard to forecast because of its 
high volatility and strong dependency on external factors. 
Our results show that the majorization criterion \cite{majorization_original}, 
which was designed as a complexity measure for random quantum circuits \cite{majorization},
is a superb performance indicator for our time series forecasting. 
That is, the quantum reservoirs with high complexity according to this criterion also exhibit better
performance in machine learning tasks. 

The performance of the QRC method has been gauged both in terms of the MAE and the MDA 
(the latter better captures the accuracy of predicting tendency changes in the data \cite{anticipating}). 
Our results indicate that the quantum reservoirs formed by gates of the G3 family 
(see Sect.~\ref{sec:QR} below) are the most suitable for this task, 
as they provide a better performance in both MAE and MDA, 
while still using only fundamental quantum gates that are easily implemented on the available NISQ devices. 
Furthermore, it is noteworthy that the optimal quantum reservoir can be implemented using only 
$\sim$100 quantum gates, which is notably less than those required for the widely used Ising model 
with the same gate set.

\section*{Acknowledgments}
The project that gave rise to these results received the support of a fellowship from 
``la Caixa'' Foundation (ID 100010434). 
The fellowship code is LCF/BQ/DR20/11790028.
This work has also been partially supported by the Spanish Ministry of Science, 
Innovation and Universities, Gobierno de Espa\~na, under Contracts 
No.\ PID2021-122711NB-C21, and by DGof Research and Technological Innovation of the Comunidad de Madrid
(Spain) under Contract No.\ IND2022/TIC-23716.

\section*{Author contribution statement}
All authors developed the idea and the theory. LD performed the calculations and analyzed the data. All authors contributed to the discussions and interpretations of the results and wrote the manuscript.

\section*{Competing financial interest statement}
The authors declare to have no competing financial and non-financial interests.

\section*{Data availability}
The raw data used for this study are publicly available from Observatorio de Precios y Mercados, 
Junta Andalucia, Spain at \CB{\url{https://www.agroprecios.com/es/precios-subasta/}} .

\bibliography{bibliography}

\end{document}